\documentclass[twocolumn,showpacs,superscriptaddress]{revtex4}
\usepackage{graphicx}
\usepackage{epstopdf}
\usepackage{amsmath}

\begin{document}
\draft
\title{Self-Assembly of Patchy Particles into Polymer Chains: A Parameter-Free Comparison between Wertheim Theory and Monte Carlo Simulation}
\author{  Francesco Sciortino} 
\affiliation{ {Dipartimento di Fisica and  INFM-CNR-SOFT, Universit\`a di Roma {\em La Sapienza}, Piazzale A. Moro 2, 00185 Roma, Italy} }
\author{ Emanuela Bianchi} 
\affiliation{ {Dipartimento di Fisica and  INFM-CNR-SMC, Universit\`a di Roma {\em La Sapienza}, Piazzale A. Moro 2, 00185 Roma, Italy} }
\author{Jack F. Douglas} 
\affiliation{{ Polymers Division, 
National Institutes of Standards and Technology, Gaithersburg,  Maryland 20899, USA}} 
   \author{Piero Tartaglia} 
\affiliation{ {Dipartimento di Fisica and  INFM-CNR-SMC, Universit\`a di Roma {\em La Sapienza}, Piazzale A. Moro 2, 00185 Roma, Italy} }

\begin{abstract}
We numerically study a simple fluid composed of particles having a hard-core repulsion, complemented by two short-ranged attractive (ÔstickyÕ) spots at the particle poles, which provides a simple model for equilibrium polymerization of linear  chains. The simplicity of the model allows for a close comparison, with no fitting parameters, between simulations and theoretical predictions based on the Wertheim perturbation theory, a unique framework for the analytic prediction of the properties of self-assembling particle systems in terms of molecular parameter and liquid state correlation functions. This theory has not been subjected to stringent tests against simulation data for ordering  across the
polymerization transition. We numerically determine many of the thermodynamic properties governing this basic form of self-assembly (energy per particle, order parameter or average fraction of particles in the associated state, average chain length, chain length distribution, average end-to-end distance of the chains, and the static structure factor) and  find that predictions of the Wertheim theory accord remarkably well with the simulation results.  
\end{abstract}


\maketitle


\section{Introduction}

      Recently, there has been great interest in exploiting self-assembly to create functional nanostructures in manufacturing, and this challenge has stimulated a great deal of experimental and theoretical activity~\cite{Manoh_03,dna,Cho_05,Yi_04,Cho_05bis,MIRKIN_96,starrnano,Glotz_04,Glotz_Solomon,zhangglotzer,doye,dnastarr,Workum_06,Stupp_93}.
      Self-assembly has been considered for over 50 years to be central to understanding structure formation in living systems and modeling and measurements of naturally occurring self-assembling systems has long been pursued in the biological sciences 
~\cite {Workum_06}. Even the term self-assembly derives from an appreciation of the capacity of viruses to spontaneously reconstitute themselves from their molecular components~\cite  {vecchio,vecchio2},
 much as in the familiar example of micelle formation by block copolymers, lipids, and other ÔsurfactantÕ molecules exhibiting amphiphilic interactions. The diversity and morphological and functional complexity of viruses, and the vast number of biological processes that form by a similar process in living systems, point to the potential of this type organizational process for manufacturing new materials. While the potential of self-assembly as a manufacturing strategy is clear, our understanding of how this process actually works is still incomplete and many of the basic principles governing this type of organization are unclear. An evolutionary (trial and error) approach to this problem is not very efficient for manufacturing. There is evidently a need for developing a first principles understanding of this phenomenon where no free parameters are involved in the theoretical description  to elucidate the fundamental principles governing self-assembly and the observables required to characterize the interactions governing thermodynamic self-assembly transitions, at least for simple model systems that can be subjected to high resolution investigation.
       
As a starting point for this type of fundamental investigation of self-assembly, we consider the problem of the equilibrium polymerization of linear polymer chains~\cite {Greer88,Greer96,Greer02},
which is arguably the simplist variety of thermally reversible particle assembly into Ôextended objectsÕ (polymers in the case of our molecular model).  To investigate this problem analytically, we exploit the Wertheim thermodynamic perturbation theory (W-TPT), which offers a systematic molecularly-based framework for calculating the thermodynamic properties of self-assembling systems, although this theory has rarely before been applied to this purpose~\cite {economou}. The Wertheim theory has been previously considered to better understand properties of associating fluids and a similar sticky sphere model to the one we consider below has been considered to determine how particle association affects critical properties associated with fluid phase separation (critical temperature and composition, binodals, critical compressibility factor, etc.~\cite {Jack_88,Busch_96,Muller},
to determine the effect of association on nucleation 
~\cite {Talan_00}
and model antigen-antibody bonding ~\cite {Busch_94}. In contrast, we are concerned here with the thermodynamic transition that accompanies the self-assembly of the particles into organized structures due to their anisotropic interactions.

The Wertheim theory is certainly not a unique theory of the thermodynamics of self-assembling particle systems. Models of equilibrium polymerization of linear, branched and compact structures have all been introduced based on the concept of an association equilibrium being established between the assembling particles and comparison of this type of theory to simulation has led to remarkably good agreement 
~\cite {rouault,Milchev98,douglasStock,Workum_05,Kindt, KindtMC,Stamb_05,Dudo_04,Workum_06}. Up to the present time, however, it has been necessary to adjust the entropy of association 
parameter in this class of theories, so that the modeling is not really fully predictive (see discussion in Sect. III where this quantity is explicitly determined from Wertheim theory).  The unique aspect of W-TPT  is that all the interaction parameters of this theory can be directly calculated from knowledge of the intermolecular potential and standard liquid state correlation functions, so that the theory is fully predictive. It has also been shown that W-TPT is formally equivalent to association-equilibrium models of self-assembly~\cite {economou}, so that the Wertheim theory also offers the prospect of being able to improve the predictive character of these other theories if the theory itself can be validated as being reliable. The Wertheim theory itself is based on a formal perturbation theory
~\cite {Werth1,Werth2,Werth3} and there are naturally questions about the accuracy that can be expected from this theory. The present paper considers a stringent test of Wertheim theory as a model of the thermodynamics of self-assembly by comparing precise numerical MC data for the thermodynamic properties of our model associating fluid to the analytic predictions of the Wertheim theory where there are no free parameters in the comparison. Notably, many of the properties that we consider have never been considered before in Wertheim theory.

\section{Two patchy sites particle model} 
\begin{figure}[t]
 \includegraphics[width=3.cm, clip=true]{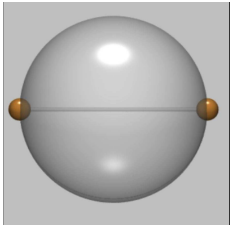}
   \caption{Pictorial representation of the model studied. Particles
   are modeled as hard-core spheres (grey large sphere of diameter $\sigma$), decorated by   two sites located on the surface along a diameter (centers
   of the   small gold 
    spheres of diameter $2 \delta$). The volume of the gold sphere outside
the grey large sphere is the bonding volume. When the site of
another particles is located within the bonding volume, the pair interaction energy is taken equal to $-u_0$.
   }
   \label{fig:m2}
\end{figure}

We focus on a system of  hard-sphere (HS) particles (of diameter $\sigma$, the unit of length) whose surface is decorated by $M=2$  identical sites  oppositely located (see Fig.~\ref{fig:m2}). 

The  interaction $V({\bf 1,2})$  between particles {\bf 1} and  {\bf 2} is 
\begin{equation}
V({\bf 1,2})=V_{HS}({\bf r_{12}})+\sum_{i=1,2}\sum_{j=1,2} V_{W}({\mathbf r}^{_{ij}}_{_{12}})
\end{equation}
where  $V_{HS}$ is
the hard-sphere potential, $V_{W}(x)$ is a square-well interaction (of depth $-u_0$ for
$x \leq  \delta$, 0 otherwise) and  ${\bf r_{12}}$ and ${\mathbf r}^{_{ij}}_{_{12}}$  are respectively the vectors joining the particle-particle centers and the site-site (on different particles) locations. 
Temperature is measured in units of the potential
depth (i.e. Boltzmann constant $k_B=1$).
Geometric considerations for a  three touching spheres configuration show that the choice of well-width $\delta=0.5(\sqrt{5-2\sqrt{3}}-1)  \approx 0.119$ guarantees that each site is engaged at most in one bond.   Hence, each particle can be form only up to two bonds and, correspondingly, the lowest  energy per particle  is $-u_0$. 

The choice of a simple square-well interaction model to describe the
association process between different particles is particularly convenient from a theoretical point of view. 
It allows for a clear definition of bonding and a clear separation of the bond free energy in an energetic and entropic contributions, being  unambiguously related to the depth of the well and to the bonding volume, respectively.

\section{Wertheim Theory}

 The  first-order Wertheim  thermodynamic perturbation 
 theory~\cite{Werth1,Werth2,Hansennew} provides an expression for the free energy of associating liquids. The Helmholtz free energy is written as  a sum of the HS reference free energy $A_{HS}$ plus a  bond contribution $A_{bond}$, which derives by a summation over certain classes of relevant graphs in the Mayer expansion~\cite{Hansennew}. In the sum, closed loops graphs are neglected.   The fundamental assumption of W-TPT  is that the conditions of steric incompatibilities are satisfied:  (i) no site can be engaged in more than one bond; (ii) no pair of particles can be double bonded.
These steric incompatibilities are satisfied in the present model thanks to the location of the two sites and the chosen $\delta$ value.
In the  formulation of Ref.~\cite  {Chap1}, for particles with two identical bonding sites, 
\begin{equation}
\label{eqn:FreeEnergyBond}
\frac{\beta A_{bond}}{N}=2 \ln{X} - X + 1
\end{equation}
Here $\beta=1/k_BT$ and  $X$  is the fraction of sites  that are not bonded. 
 $X$ is calculated  from the mass-action equation
\begin{equation}\label{eqn:Xa}
X=\frac{1}{1 + \displaystyle 2 \rho  X \Delta}
\end{equation}
where $\rho=N/V$ is the particle number density and $\Delta$ is defined by
\begin{equation}\label{eqn:Deltabis}
\Delta= 4 \pi \displaystyle \int{g_{HS}(r_{12})\langle f(12)\rangle _{\omega_{1},\omega_{2}} r_{12}^{2} d r_{12}}
\end{equation}
 Here $g_{HS}(12)$ is the reference HS fluid pair correlation function, the Mayer $f$-function is $f(12)=exp(-V_{W}({\mathbf r}^{_{11}}_{_{12}})/k_{B}T)-1$, and   $\langle f(12)\rangle _{\omega_{1},\omega_{2}}$~\cite  {Werth5}  represents an 
 angle-average over all orientations of particles 1 and 2 at fixed relative distance $r_{12}$.  Since all bonding sites are identical (same depth and width of the square-well interaction), $\Delta$ refers to a single site-site interaction.  The number of attractive sites on each particle is encoded in the factor $2$ in front of $\Delta$ in  Eq.~\ref{eqn:Xa}. In the W-TPT, the resulting free energy is insensitive to the location of the attractive sites, i.e. to the bonding geometry of the particle.
Note that the angle averaged Mayer $f$-function coincides with
the bonding interaction contribution to the virial coefficient. At low $T$ (i.e. $\beta u >> 1$)
the hard-core contribution to the virial becomes negligible as compared to the bonding component.  In this limit, it is also possible
to assume $\exp(\beta u_0) - 1 \approx \exp(\beta u_0) $, so that
the  averaged Mayer $f$-function can be approximated with the virial as well as with the integral of the Boltzmann factor over the bond volume\cite{Searco}.

 For a site-site  square-well interaction, the Mayer function can be calculated as ~\cite  {Werth5}
\begin{equation}
\langle f(12)\rangle _{\omega_{1},\omega_{2}} =  \left [
\exp(\beta u_0)-1 \right ] S(r) 
\end{equation}
where 
\begin{equation}
S(r)=\frac{(\delta + \sigma -r)^2  (2 \delta - \sigma +r )}{ 6 \sigma^2 r}
\end{equation}
is the fraction of solid angle available to bonding when two particles are located at relative center-to-center distance $r$.  
%
Thus the evaluation of $\Delta$  
requires only an expression for
$g_{HS}(r_{12})$  in the range where bonding occurs ($\sigma<r<\sigma+\delta$). 
We have used the linear approximation~\cite  {Nez_90} 
\begin{equation}
g_{HS}(r)= \frac{1-0.5 \phi}{(1- \phi)^3}-\frac{9}{2}\frac{\phi (1+\phi)}{(1- \phi)^3} \left[\frac{r-\sigma}{\sigma}\right]
\label{eqn:ghsr}
\end{equation}
(where $\phi=\frac{\pi}{6} \sigma^3 \rho$) which provides the correct Carnahan-Starling~\cite  {CS_69} value at contact. This gives
\begin{eqnarray}
\label{eq:deltamodel}
\Delta= 
\frac{V_b (e^{\beta u_0} - 1)}{ (1-\phi )^3}  \times \\ \nonumber
\left[1-\frac{5}{2}\frac{\left(3+8 \delta +3 \delta ^2\right)  }{ (15+4 \delta ) }\phi -\frac{3}{2}\frac{\left(12 \delta +5 \delta
^2\right) }{(15+4 \delta ) }\phi ^2\right] 
\end{eqnarray}
where we have defined the spherically averaged bonding volume $V_b \equiv 
 4 \pi \int_\sigma^{\sigma+\delta} S(r) r^2 dr =\pi \delta^4 (15+4 \delta) /30
 $. 
For the specific value of $\delta$  studied here, 
$V_b = 0.000332285~\sigma^3$.
At low $\phi$, $g_{HS}(r)$ tends to the ideal gas limit value $g_{HS}(r) \approx 1$.  In this limit
\begin{eqnarray}
\Delta=V_b (e^{\beta u_0} - 1).
\label{eq:deltaig}
\end{eqnarray}

Eq.~(\ref{eqn:Xa}) can be easily solved, providing the 
$T$ and $\rho$ dependence of $X$ as
\begin{equation}
X=\frac{2}{1+\sqrt{1+8 \rho \Delta}},
\label{eq:X}
\end{equation}
which has a low $T$ limit  $X \approx \sqrt{2 \rho \Delta}$.

In the more transparent chemical equilibrium form, Eq.~(\ref{eq:X})
can be written as
\begin{equation}
\frac{1-X}{X^2} = 2 \rho \Delta = \rho K_b.
\label{eq:kb}
\end{equation}
The last expression shows that, within Wertheim theory,
bonding can be seen as a chemical reaction between two
unreacted sites forming a bonded pair.  In this language
the quantity $K_b \equiv 2 \Delta$ plays the role of equilibrium constant  (in unit of inverse concentration).
Writing $\rho K_b= \exp{\{-\beta (\Delta  U_b-T\Delta S_b)\}}$ (introducing the 
energy and entropy change in the bond process), it is possible to
provide precise expressions for $\Delta U_b$ and $\Delta S_b$ within the Wertheim theory.  Specifically,   when $e^{\beta u_0}>>1$ (a very minor approximation since aggregation requires $T << u_0$ to be effective) it is possible to identify 
\begin{eqnarray}
\Delta U_b=-u_0 \\
\Delta S_b=  \ln \left [8 \pi \rho \displaystyle \int_\sigma^{\sigma+\delta} g_{HS}(r) S(r) r^2 dr \right ]
\end{eqnarray}
(note that $\Delta S_b$ is in dimensionless entropy units, since Boltzmann's constant is taken to equal 1.).
If $g_{HS}\approx 1$, $\Delta S_b=  \ln(2N V_b/V)$.
 Hence the change in energy is given by the bond energy, while the change in entropy is essentially provided by the logarithm of the ratio between the bonding volume and the  volume per site ($V/2N)$.

\section{Cluster size distributions and association properties}

It is  
interesting  to discuss the prediction of the Wertheim theory in term of  clusters of physically bonded particles~\cite  {hill,coniglio}.
 In the case of square-well interactions (differently from a continuous potential) we can define a bond between two particles 
 unambiguously. 
Evidently, when there is a bond the interaction energy equals - $u_0$.

 To make the discussion more transparent, we can define 
$p_b \equiv 1-X$ as the probability that an arbitrary site is bonded.  
It is thus easy to convince oneself that the number density of
monomers 
is
$\rho_1=\rho (1-p_b)^2= \rho X^2$, since both sites must be unbonded~\cite  {economou}. Similarly a chain of $l$ particles has a number density $\rho_l$
equal to 
\begin{equation}
\rho_l =\rho (1-p_b)^2 p_b^{(l-1)}= \rho X^2 (1-X)^{(l-1)}
\label{eq:csd}
\end{equation}

since one site of the first and 
one 
of the last particle in the chain must be unbonded and $l-1$ bonds link  the $l$ particles.\\

Once the cluster size distribution of chains is known,
it is possible to calculate the average chain length $L$ as
the ratio between the total number density and the  number  density of chains in the  system, (i.e., as the ratio between the first  $\langle l^1 \rangle$ and the
zero $\langle l^0\rangle$ moments  of the $\rho_l$ distribution)
\begin{equation}
L \equiv \frac{\sum_{l=1}^{\infty} l \rho_l}{\sum_l \rho_l}=\frac{1}{X}
\end{equation}
where we have substituted, by summing the geometric series  over all chain lengths,
$\sum_{l=1}^{\infty} l  \rho_l =\rho$ and  $\sum_{l=1}^{\infty}  \rho_l  = \rho X$.
Thus, using Eq.~(\ref{eq:X}), 
\begin{equation}
L=\frac{1+\sqrt{1+8 \rho \Delta}}{2}
\label{eq:l}
\end{equation}
At low $T$, $L \approx \sqrt{2 \rho \Delta}$  and hence $L$ grows in density  as $\sqrt{\rho}$ (if the density dependence of $\Delta$ can be neglected~\cite {candau}) and in $T$ as $L \sim exp(\beta u_0/2)$.

The potential energy of the system  coincides with the number of bonds (times $-u_0$). 
Hence, the energy per particle $E/N$ is
\begin{equation}
E/N=-u_0 \frac{\sum_{l=1}^{\infty} (l-1)  \rho_l }{\sum_{l=1}^{\infty} l  \rho_l }=-u_0 (1-X)=-u_0 p_b
\label{eq:e}
\end{equation}
The same result is of course obtained calculating
$E/N$ as $\frac{\partial (\beta A_{bond}/N)}{\partial \beta}$ where ($\beta A_{bond}/N) $ is given by Eq.~(\ref{eqn:FreeEnergyBond}).
The energy approaches its ground state value ($E_{gs}/N=-u_0$)
as 
\begin{equation}
\frac{E-E_{gs}}{N}= u_0 X  \approx  u_0 (2 \rho \Delta)^{-\frac{1}{2}}
\end{equation}
i.e. with an Arrhenius law with activation energy $u_0/2$ in the low $T$ limit, a signature of independent bonding sites~\cite  {Zacca1,Zacca2,Moreno_05}.

From Eq.~(\ref{eq:e}) it is possible to calculate the (constant volume)
specific heat $C_V$ as
\begin{equation}
C_V= C \frac{X^2  e^{\beta u_0 }\rho} {\sqrt{ 1 + 8 \rho \Delta } T^2}
\label{eq:cv}
\end{equation}
where $C = 8 \pi u_0^2 \displaystyle \int_\sigma^{\sigma+\delta} g_{HS}(r) S(r) r^2 dr $.
In the  low $T$ and 
$\rho$ regime, the specific heat becomes 
 $C_V \approx X/2T^2$.
\begin{figure}[h]
\centerline {\includegraphics[width=3.5in]{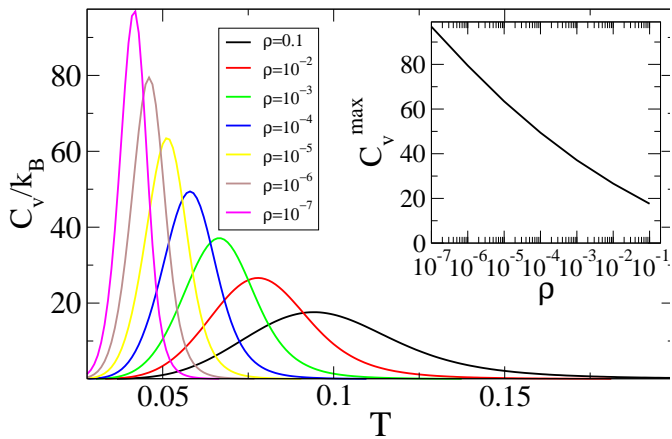}}
\caption{W-TPT predictions for the $T$-dependence of the specific heat  $C_V$ for different values of densities $\rho$ (Eq.~\ref{eq:cv}). The inset shows the value of the specific heat at the maximum $C_V^{max}$. }
\label{fig:cv}
\end{figure}
At each $\rho$, the  specific heat shows a maximum at a finite $T$ (see Fig.~\ref{fig:cv}), which define a lines of specific heat extrema in the   $T-\rho$ plane. The location of the maximum in the specific heat has been utilized to estimate the polymerization 
temperature~\cite{Greer88,Greer96,Greer02,Milchev98,douglas,Dudo_03}

\begin{figure}[h]
\centerline {\includegraphics[width=3.5in]{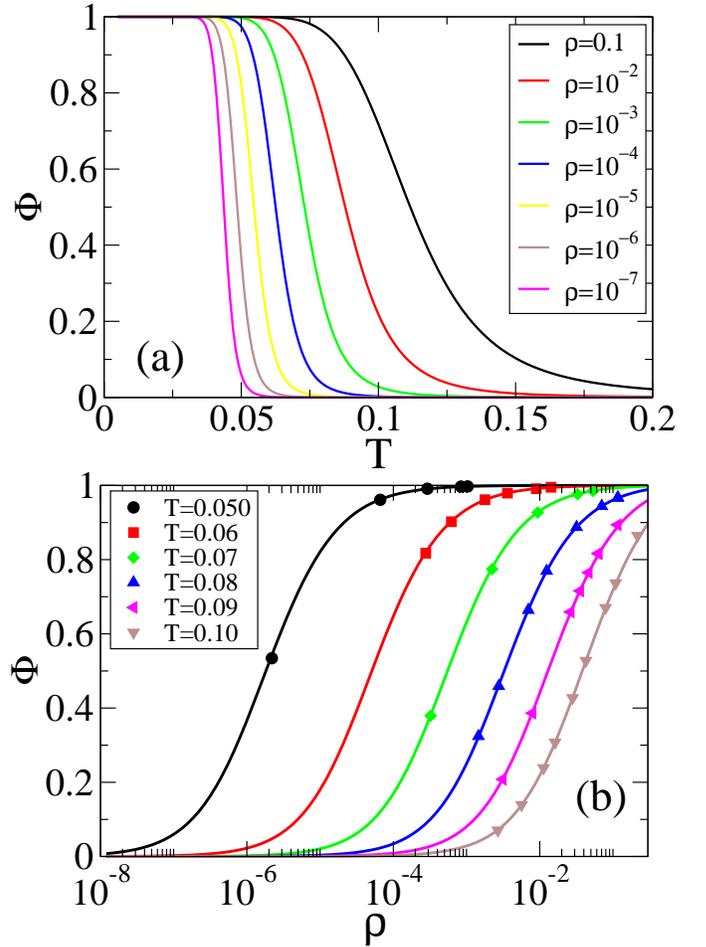}}
\caption{Extent of polymerization $\Phi$ vs. temperature  $T$ (a) and density $\rho$ (b). Symbols are GC simulation data.}
\label{fig:Phi}
\end{figure}

Within the theory, it is also possible to evaluate the extent of
polymerization $\Phi$,  defined as the fraction of particles connected in chains (chain length 
larger than one), i.e.
\begin{equation}
\Phi=\frac{\sum_{l=2}^{\infty} l  \rho_l } {\sum_{l=1}^{\infty} l \rho_l }= 1- \frac{\rho_1}{\rho}=1- X^2,
\label{eq:Phi}
\end{equation}
\noindent
where we have used Eq.~\ref{eq:csd}.  
$\Phi$ plays the role of
order parameter for the polymerization transition. The density and temperature dependence of $\Phi$ are shown in Fig.~\ref{fig:Phi}.
The cross-over from the
monomeric state at high $T$ ($\Phi \approx 0$) to the
polymeric thermodynamic state at low $T$ ($\Phi \approx 1$) takes place in a
progressively smaller $T$-window 
 on decreasing $\rho$

Indeed, for each $\rho$, a transition temperature $T_{\Phi}$ can be defined as the {\it{inflection point}} of $\Phi$ as function of $T$~\cite{Greer88,Greer96,Greer02,douglas,Dudo_03}. The locus $T_{\Phi}(\rho)$ provides an estimate of the polymerization line in the phase diagram.  The location of the inflection point of the energy $E$ and of $\Phi$ are different, since $dE/dT \sim dX/dT$ (Eq.~\ref{eq:e}), while $d\Phi/dT \sim -X dX/dT$ (Eq.~\ref{eq:Phi}).
In other models of equilibrium polymerization, incorporating thermal activation or chemical initiation $T_{\Phi}$ and $T_{C_V^{max}}$ coincide~\cite{douglas,Transref}.

Another estimate of the transition line of this rounded thermodynamic transition
can be defined as the locus in the $T-\rho$ plane at which $\Phi=0.5$, i.e. half of the particles are in chain form (the analog of the critical micelle concentration~\cite  {jones}), i.e. $\rho_1(T_{1/2},\rho)=\rho/2$. The corresponding temperature $T_{1/2}$ is then given by the solution of the equation  $X=\sqrt{0.5}$, or equivalently
\begin{equation}
\rho \Delta=1-\sqrt{0.5}
\label{eq:t1/2}
\end{equation}
In the present model, the $\Phi=0.5$ locus is a  line at constant  $p_b$,   corresponding to a
constant value of the product $\rho~[\exp(\beta u_0)-1]$.

\begin{figure}[h]
\includegraphics[width=8cm, clip=true]{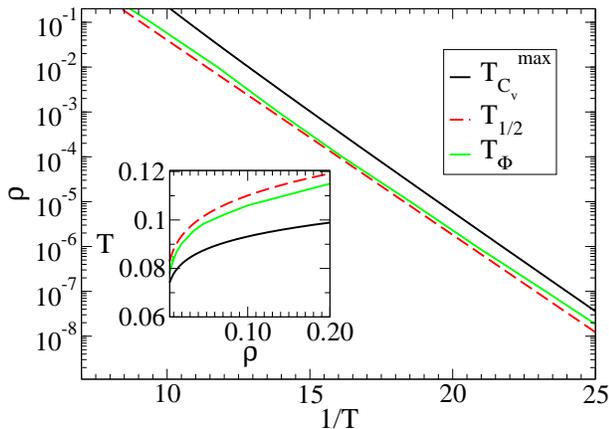}
\caption{Specific heat maxima and polymerization transition lines, as predicted by the Wertheim theory, in the $T-\rho$ plane. The inset shows the linear scale. Observe the similarity of the data in Fig.~2-4 to that of Fig.~4 - 6 in Ref.~\cite  {Workum_05} corresponding to equilibrium polymerization in the Stockmayer fluid, a Lennard-Jones particle with a superimposed point dipole. 
}
\label{fig:prediction}
\end{figure}

To provide a global view of the polymerization transition, we show
in Fig.~\ref{fig:prediction} the Wertheim theory predictions for the
specific heat maximum and  polymerization transition lines. Curves becomes progressively more and more similar on cooling. As clearly shown by the simple expression for $T_{1/2}$ (Eq.~\ref{eq:t1/2})  the quantity $ \rho \Delta$ is constant, which implies that $\ln \rho \sim 1/T$. This behavior is also approximatively found for the loci defined by the 
inflection point of $\Phi$ and $E$, as shown in Fig.~\ref{fig:prediction}.

\begin{figure}[h]
\includegraphics[width=8cm, clip=true]{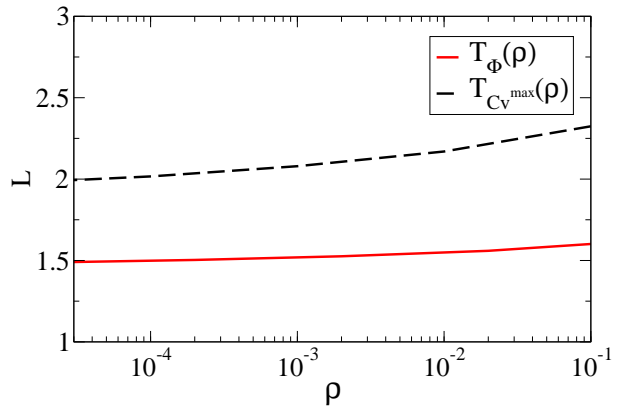}
\caption{Value of the average chain length along the polymerization lines $T_{\Phi}(\rho)$ and $T_{C_V^{max}}(\rho)$, according to the Wertheim theory. }
\label{fig:lat}
\end{figure}

It is interesting to evaluate the value of  average degree of polymerization $L$ along the 
 polymerization transition line. This information helps estimating  the polymerization transition itself, but it is also  relevant for the recently proposed analogies between polymerizing systems and  glass-forming liquids~\cite  {jackglass}.  As shown in Fig.~\ref{fig:lat} the transition takes place for $L \approx 2$, consistent with previous findings (compare with the inset of Fig. 7 
in Ref.~\cite  {Workum_05} for the Stockmayer fluid).  The fact that, at  $T_{\phi}$, $L \approx 2$
can provide a way to locate the transition temperature when experimental (or numerical)  data are noisy~\cite {Stamb_05}.

A final relevant consideration concerns the expression for the bonding free energy.  Under the assumption of an ideal state of chains, the
system free energy $A$ can be written as
\begin{equation}
\beta A/V = \sum_{l=1}^{\infty}  \rho_l  \left [ \ln{ \rho_l }-1 + (l-1) \ln{K_b} \right]
\end{equation}
which accounts for the translational entropy of the clusters as well as for the bond free energy $\ln{K_b}$ which is assumed to be linear in the number of bonds. At high $T$ only monomers are present, $ \rho_l =\rho \delta_{l1}$ and hence
$\beta A/V = \rho  [\ln{\rho}-1]$.  The bonding part of the free energy $\beta A_{bond}$ can thus be evaluated as difference of $\beta A$ and the corresponding high $T$ limit, i.e.
\begin{equation}
\beta A_{bond}/V= \sum_{l=1}^{\infty}  \rho_l  \left [ \ln{ \rho_l }-1 + (l-1) \ln{K_b} \right] -  \rho  [\ln{\rho}-1]
\end{equation}
Inserting the Wertheim cluster size distribution (Eq.~\ref{eq:csd}),
summing over all cluster size and using the definition $K_b \equiv 2 \Delta$  (Eq.~\ref{eq:kb}), one exactly recovers  the Wertheim
bonding free energy (Eq.~\ref{eqn:FreeEnergyBond}).
The Wertheim theory can thus be considered as a mean-field theory of chain associations, with the hard-sphere free energy as its high-$T$ limit. Differently from other mean-field approaches~\cite  {candau,Searprl,douglas,Kindt}, the theory provides also a well-defined prescription for calculating the equilibrium constant, which partially account for the structure of the reference hard-sphere fluid via the $g_{HS}(r)$ contribution in $K_b$. It is this special feature that makes the theory fully predictive.

\section{Monte Carlo Simulation}

We have performed numerical simulations of the model
in the grand-canonical (GC) ensemble~\cite  {frenkelsmith}  for several values of $T$ and of  activities to evaluate the
structural properties of the system as a function of density and
temperature.  
We have performed two types of grand canonical simulations: particle and chain insertion/removal.

 The first method (particle) is a classical GC MC simulation where monomers are individually added to or eliminated from the system with insertion 
 and removal probabilities 
  given by
\begin{eqnarray}
P_{insertion}& = & min(1, \frac{ z   V}{N+1} e^{-\beta \Delta E} ) \\
P_{removal}& = & min(1,\frac{N}{V}   \frac{e^{\beta \Delta E}}{ z}),
\end{eqnarray}
where $N$ is the number of monomers, $\Delta E$ is the change in the system energy upon insertion (or removal) and $z$ is the
chosen activity. We have simulated for about two million MC steps, where a MC step has been defined as 50000 attempts to move  a particle  and 100 attempts to insert or delete a particle.  A  move is defined as a displacement  in each direction of a random quantity distributed uniformly between $\pm ~0.05~\sigma$ and a rotation around a random axis of random angle distributed uniformly 
between $\pm 0.1$ radiant.  The box size has been fixed to 50 $\sigma$. With this type of simulations we have studied three temperatures (T~=~0.08,~0.09 and 0.1) and several densities
ranging from 0.001 up to 0.2 (corresponding to number $N$ of particles ranging from  $N \approx 1000$ to  $N \approx 20000$.

The second method is a grand-canonical simulations  where we had and remove chains of particles (see for example Ref.~\cite{KindtMC}).  The insertion 
 and removal probabilities 
 for a chain of length $l$ are:

\begin{eqnarray}
P_{insertion}& = & min(1, \frac{ z^l  e^{-(l-1) \beta  f_b} V}{N_i+1} e^{-\beta \Delta E} ) \\
P_{removal}& = & min(1,\frac{N_l}{V}   \frac{e^{\beta \Delta E}}{ z^l  e^{-(l-1) \beta  f_b}})
\end{eqnarray}
where $N_l$ is the number of chains of size $l$, $\Delta E$ is the change in energy and  $\beta f_b$ is the $\ln$ of the integral of the Boltzmann factor over the  bond volume, i.e., 
\begin{equation}
\beta f_b =\beta u_0 -\ln[ 2 V_b ] 
\end{equation}
The  activity of a chain of $l$ particles is thus written as
$z_l=z^l  e^{-(l-1) \beta  f_b}$, or equivalently in term of chemical potential $\mu_l=l \mu - (l-1) f_b$, where $\mu$ is the monomer chemical potential.  
Using the chain insertion algorithm,  we have followed the system for about $10^{5}$ MC steps, where a MC step has now been defined as 50000 attempts to move  a particle (as in the previous type) and 100 attempts to insert or delete a chain of randomly selected length.  The geometry of the chain to be inserted is also randomly selected.  The box size was varied from 50 to 400 $\sigma$, according to density and temperature, to guarantee that the longest chain in the system was always shorter than the box size.  At the lowest $T$,   $N  \approx 50000$. Using chain moves we have been able to equilibrate and study densities ranging from $10^{-6}$ up to $0.2$ for four 
low temperatures (T~=~0.05, 0.055, 0.6 and 0.7)  where chains of length up to 400 monomers are observed.  We have also studied the same three temperatures (T~=~0.08, 0.09 and 0.1) examined  with the particle insertion/removal method to compare the two MC approaches. 

As a result of the long simulations performed  (about three months of CPU time for each state point) resulting  average quantities 
calculated from the MC data ($L$, $\Phi$, $E/N$) are affected by less than 
3~\% relative error. Chain length distributions  $\rho_l$ (whose signal  covers up to six order of magnitude) are affected by
an error proportional to $|log(\rho_l)$ which progressively increases on decreasing $\rho_l$, reaching 70\% at the smallest reported $\rho_l$ values.

\section{
Simulation results}

\begin{figure}[t]
\includegraphics[width=2in]{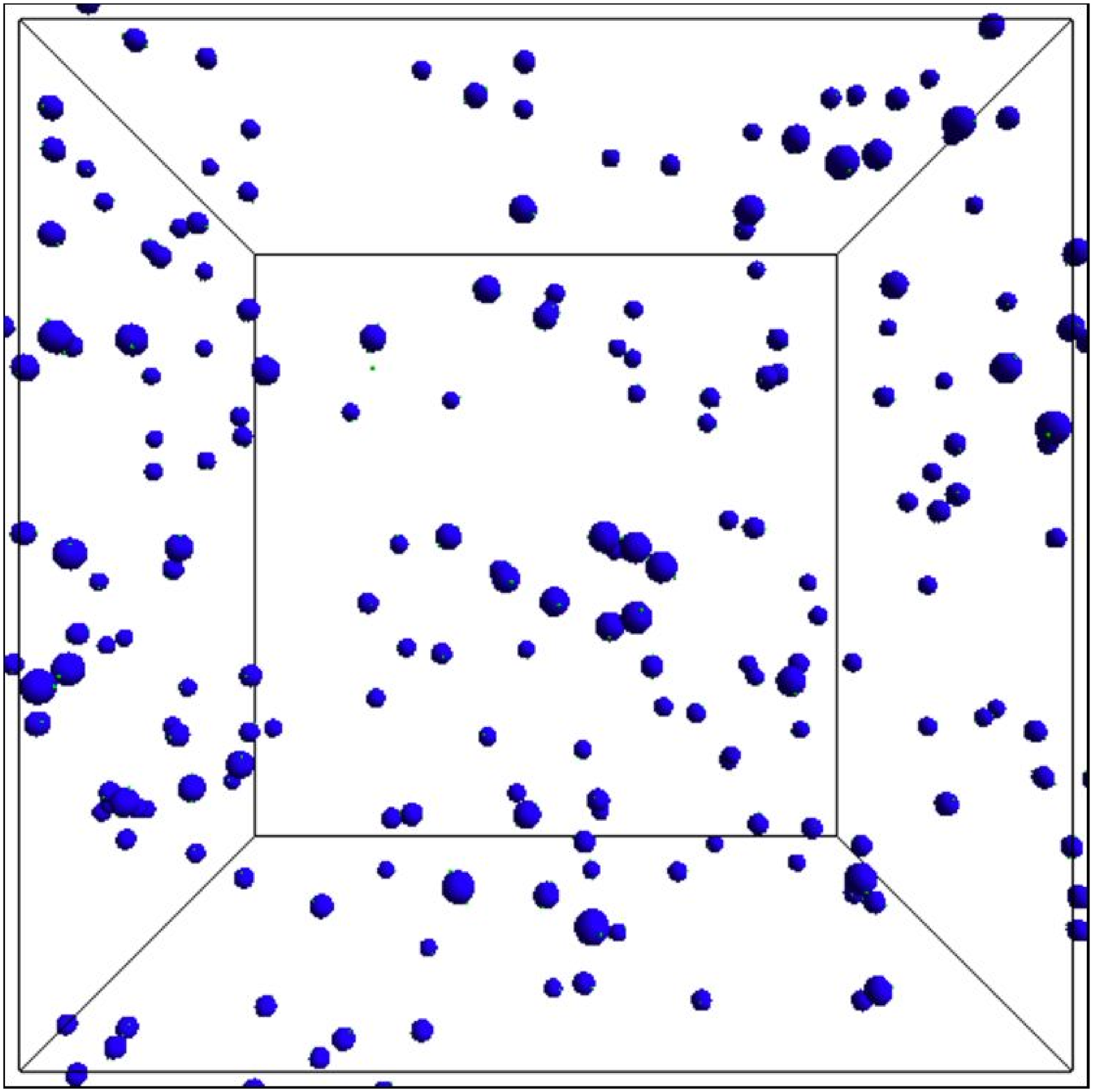}
\includegraphics[width=2in]{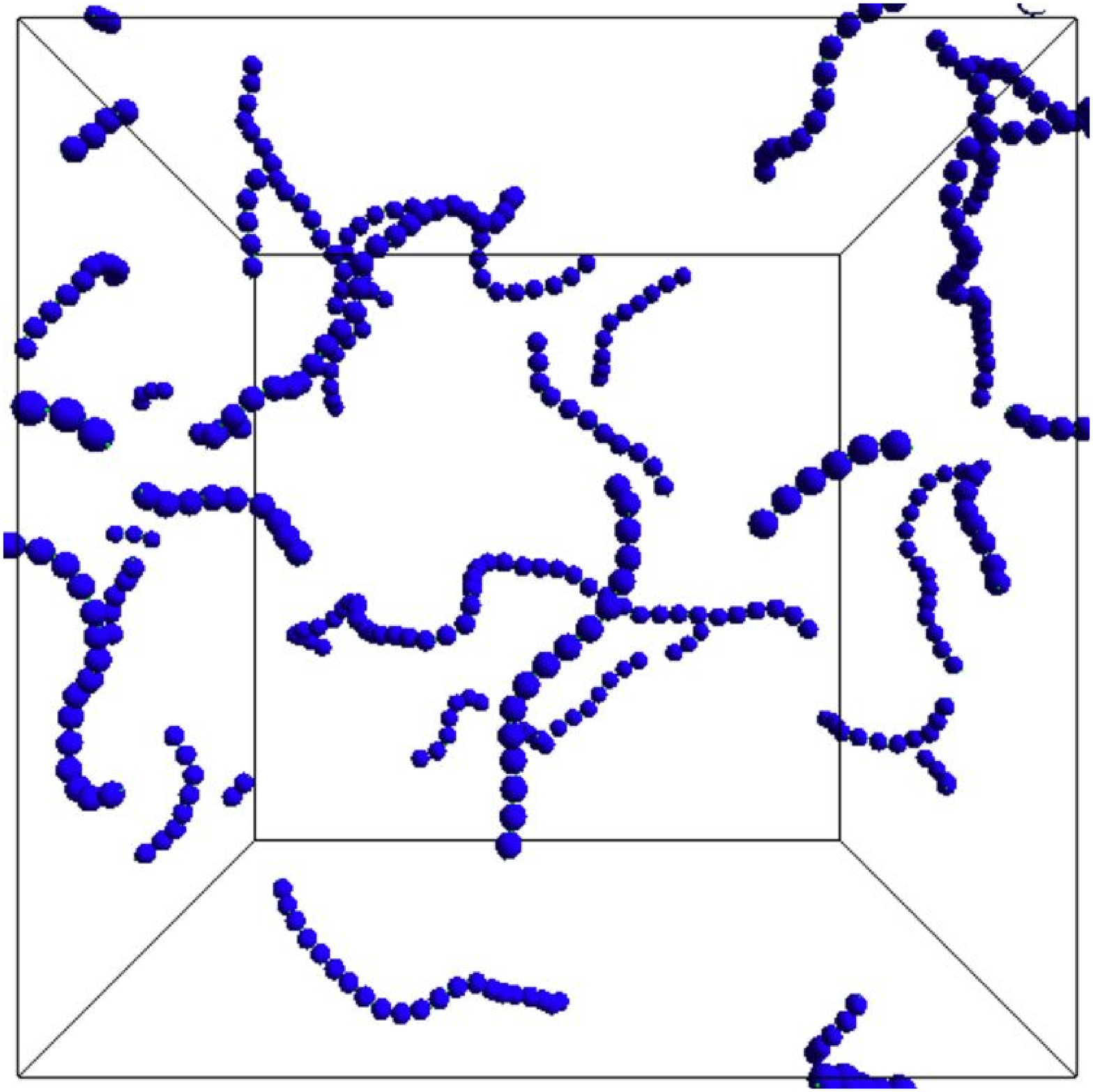}
\caption{Snapshots of a fraction of the simulated system at $\rho \approx 0.0035$ and $T=0.10$ and $T=0.055 $. At this density, the polymerization transition is located at
$T_{\Phi} \approx 0.08$.
The length of the shown box edge is about $30 \sigma$. }
 \label{fig:snapshot}
\end{figure}

\begin{figure}[t]
 \includegraphics[width=9cm, clip=true]{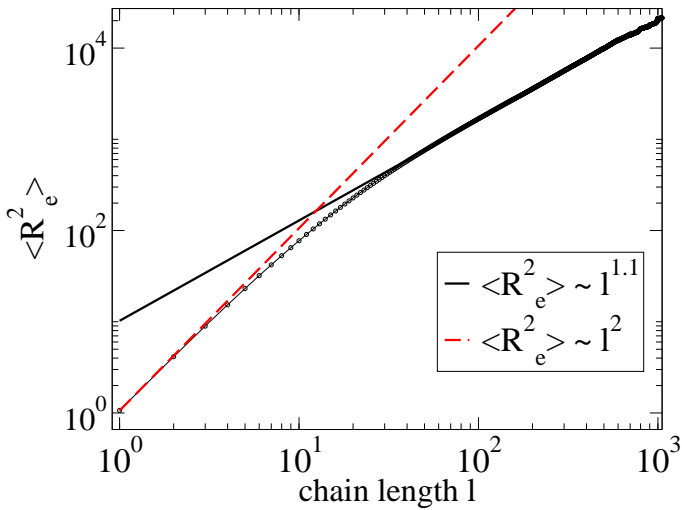}
   \caption{Mean squared end-to-end distance  $<R^2_e>$ for isolated chains  of length $l$ built according to the model studied in this article.  }
   \label{fig:endtoend}
\end{figure}

 First, we begin by showing in Fig.~\ref{fig:snapshot} representative particle configurations both above and below the polymerization transition temperature, $T_{\phi}$. Evidently, the particles are dispersed as a  gas of monomers and as a gas of  chains above and below this characteristic temperature.
The low-$T$ configurations is composed by semi-flexible chains, with no rings. Indeed,  the short interaction range $\delta$  introduces a significant stiffness 
in the chain and a persistence length 
extending over several monomers. 
To quantify the linearity of the chains for the present model we
show in  Fig.~\ref{fig:endtoend}  the chain end-to-end squared distance $<R^2_e>$ for isolated chains of chain length up to $l \sim O(10^3)$.  Single chains are generated by progressively adding monomers to a pre-existing chain in a bonding configurations, after checking the possible overlap with all pre-existing monomers.  Since the bond interaction is a well, all points
in the bond-volume have the same a-priori probability. As shown in
Fig.~\ref{fig:endtoend},
the end-to-end chain distance scales as a power-law ($<R^2_e>\sim l^{2\nu}$)  both at
small and large $l$ values, with a crossing between the two
behaviors  around  $l\simeq O(10)$. 
At small $l$, 
the chain is persistent in form and thus is rod-like. For larger $l$, the  best-fit with a
power law suggests  an apparent exponent
 $ 2\nu \approx 1.1$, that is expected to evolve --- for very long chains --- toward the self-avoiding value 
 $2 \nu \approx 1.18$~\cite  {LeGui,DouglasA}.
We recall that  $2\nu =6/5$ in the Flory mean field prediction~\cite  {flory} and  $2\nu=1$  in the simple random walk model.

\begin{figure}[h]
\includegraphics[width=9cm, clip=true]{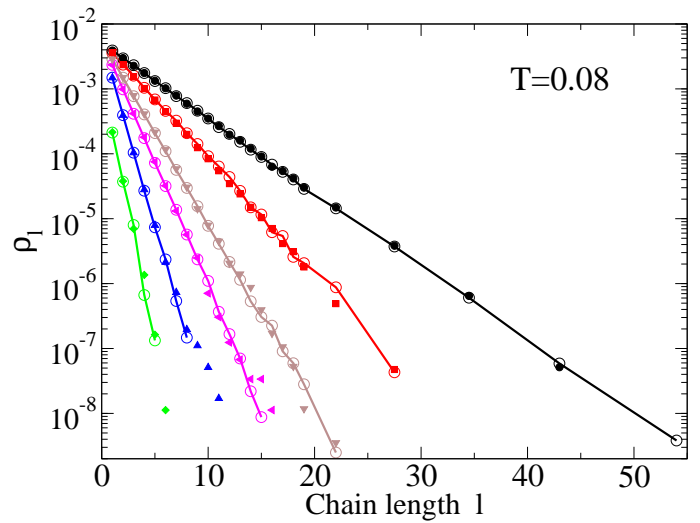}
   \caption{Comparison of the chain length distributions $\rho_l$ at $T=0.08$  obtained independently with the monomer (full symbols) and chain (lines connecting open symbols)  grand-canonical simulations for several values of the monomer  activity $z$.
From left to right, the  activity values are: $~10^{-3},1.5~\times~10^{-3},2.4~\times~10^{-3},3~\times~10^{-3},5~\times~10^{-3},6~\times~10^{-3}$.    Similar agreement is also found at the other temperatures. }
   \label{fig:twomethods}
\end{figure}

To provide evidence that the chain GC MC simulation provides the correct sampling of the configurations (and hence that the  activity of the chain of length $l$ is correctly assigned) we compare in Fig.~\ref{fig:twomethods} the chain length densities $\rho_l$  calculated with the two methods at  $T=0.08$.  The distributions calculated with the two different methods are identical. Similar agreement is also found at $T = 0.09$ and $T = 0.1$.  This strengthens the possibility of using the chain MC method, which does not significantly suffer from the slow equilibration process associated to the  increase of the Boltzmann factor $\exp{(\beta u_0)}$ on cooling. All the following data are based on chain GC-MC simulations.

\begin{figure}[h]
\includegraphics[width=9cm, clip=true]{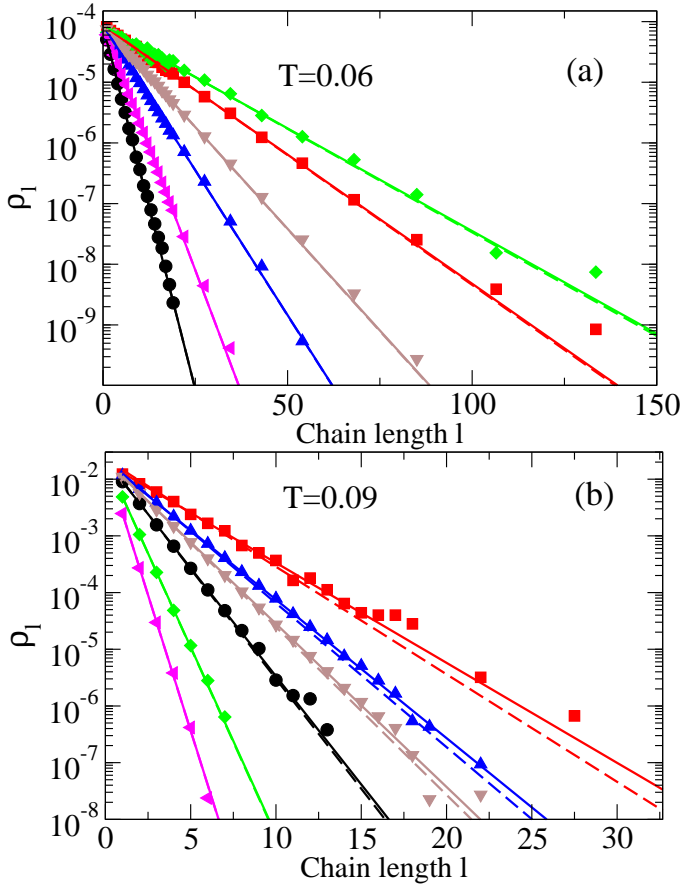}
\caption{Chain length densities  $\rho_l$  for several activity values at two different temperatures, $T=0.06$ and $T=0.09$. Symbols are simulation data. Lines are Wertheim theory predictions: dashed lines assume $g_{HS}=1$ (Eq.~\ref{eq:deltaig} for $\Delta$), while full lines are based on the full radial dependence of $g_{HS}$ (Eq.~\ref{eq:deltamodel} for $\Delta$). At the  lowest $T$ (and average densities), the ideal gas approximation is sufficient to model the Monte Carlo data.}
\label{fig:csd}
\end{figure}

We next compare the chain length distributions calculated using the chain MC method with the predictions of the Wertheim theory.
We compare the simulation data with two different approximation:
in the first one we choose the ideal gas as reference state, i.e. we approximate the reference radial distribution function with one. In the more realistic approximation, we use the 
 small $r$ expansion of the hard-sphere radial distribution function (see Eq.~\ref{eqn:ghsr}).  Comparison between simulation data and theoretical predictions (
 note there are no fitting parameters) is reported in Fig.~\ref{fig:csd} for two different temperatures. At low densities
 (sampled at low $T$)  the approximation $g_{HS}\approx 1$ is already sufficient to properly describe $ \rho_l $. At higher densities (sampled at higher $T$), the full theory is requested
to satisfactory predict the  chain length distributions.

\begin{figure}[h]
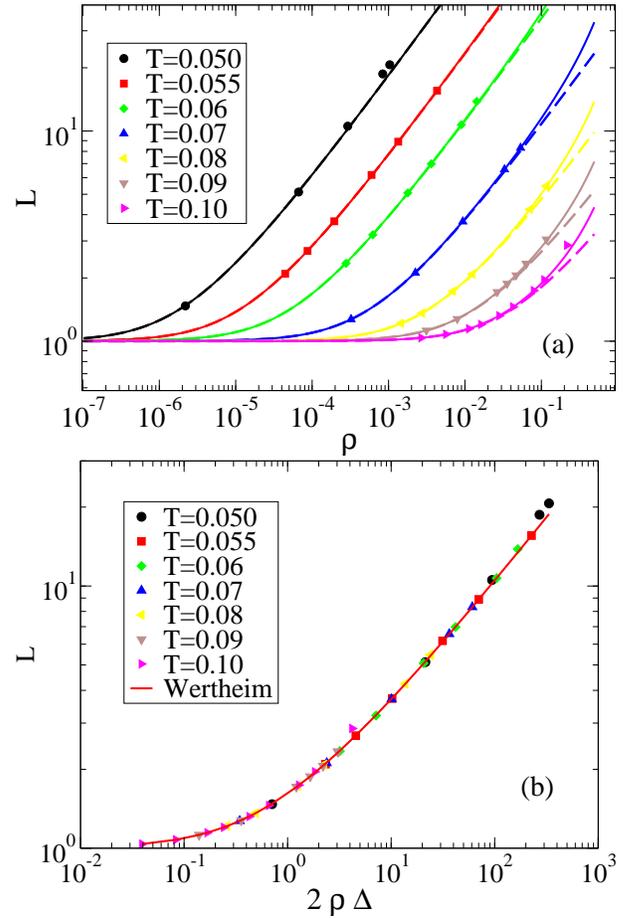

\includegraphics[width=8cm, clip=true]{lmed.eps}
\includegraphics[width=8cm, clip=true]{lscaled.eps}
\caption{Average chain length as a function of the density for all studied temperatures. (a): Lines are the Wertheim theory predictions: dashed lines assume $g_{HS}=1$ (Eq.~\ref{eq:deltaig} for $\Delta$), while full lines are based on the full radial dependence of $g_{HS}$(Eq.~\ref{eq:deltamodel} for $\Delta$). (b): Scaled representation of 
$L$ vs. $2 \rho \Delta$. 
Symbols are simulation data. The line is the function
$L(x)=\frac{1+\sqrt{1+2 x}}{2}$ (See Eq.~\ref{eq:l})
}
\label{fig:lmed}
\end{figure}

Fig.~\ref{fig:lmed}(a) compares the Wertheim theory predictions for the average chain length with the corresponding simulation results.
 In the entire investigated $\rho$ and $T$ range, the Wertheim theory provides an accurate description of the equilibrium polymerization process. The limiting growth law in $\sqrt{\rho}$ is clearly visible at the lowest temperatures. At the highest  
temperatures, it is possible to access the region of larger densities 
($\rho \sim 0.1$) where the presence of other chains can not be
neglected any longer and $\Delta$ becomes $\rho$ dependent.
In this limit, the growth low $L \sim \sqrt{\rho}$ is  no longer obeyed~\cite {candau}. As a further check on Wertheim theory, we collapse all the $L$ data
to the universal functional
 form predicted by Wertheim theory using the scaling variable
$2 \Delta \rho$ (Fig.~\ref{fig:lmed}(b))
 
\begin{figure}[h]
\includegraphics[width=8cm, clip=true]{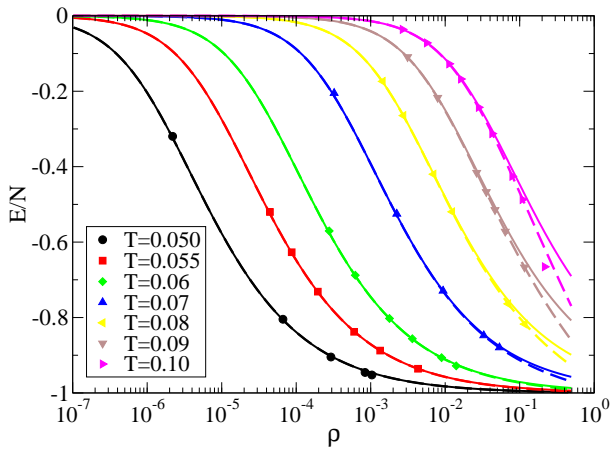}
\caption{Energy per particle in unit of $u_0$ as  a function of the density for all studied temperatures. 
Symbols are simulations data while lines are the Wertheim theory predictions: dashed lines assume $g_{HS}=1$ (Eq.~\ref{eq:deltaig} for $\Delta$), while full lines are based on the full radial dependence of $g_{HS}$ (Eq.~\ref{eq:deltamodel} for $\Delta$) }
\label{fig:Emed}
\end{figure}

As an ulterior confirmation of the predictive capabilities of Wertheim theory, we report  a comparison 
between  simulations and  theory for 
the 
density dependent of the extent of polymerization $\Phi$  
(Fig.~\ref{fig:Phi})
and for 
  the energy per particle (Fig.~\ref{fig:Emed}). Both figures clearly shows a 
  excellent agreement between the simulated and the predicted  $\rho$ dependence of  $\Phi$ and $E$ at all $T$ investigated. 

\begin{figure}[h]
\includegraphics[width=8cm, clip=true]{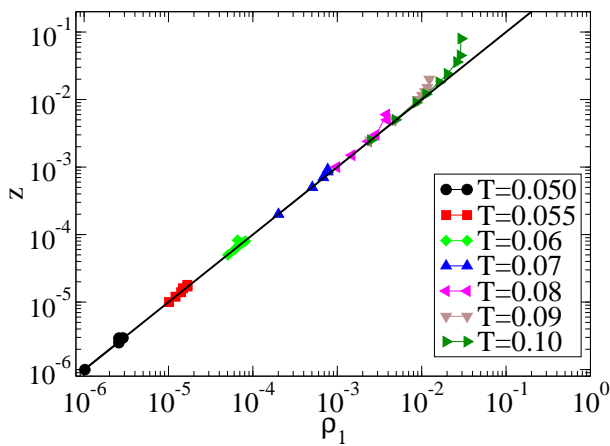}
\caption{Relation between the  activity $z$ and the monomer density $\rho_1$  for all investigated temperatures. Line indicates the ideal gas law  $z=\rho_1$.   }
\label{fig:activity}
\end{figure}

As a test of the validity of the approach of the polymerizing system as an ideal gas of equilibrium chains,  we compare the  monomer activity $z$ and the monomer density $\rho_1$ in Fig.~\ref{fig:activity}. Indeed
the activity of a cluster of size $l$ coincides with $ \rho_l$, which is consistent with ideal gas scaling. In particular, the activity of the single particle (the input in the MC grand-canonical simulation) can be compared with the resulting density of monomers $\rho_1$.  Data in Fig.~\ref{fig:activity} shows that the ideal-gas
law is well obeyed at low $T$ and $\rho$, confirming that, in the investigated range, 
the system can be visualized as an ideal gas mixture of chains of 
different lengths, distributed according to Eq.~\ref{eq:csd}.

 The existence of a large $T$-$\rho$ window where an ideal mixture of chains provides a satisfactory representation of the system suggests that, in this window, correlations between different chains can be neglected. In this limit, the structure of the system should be provided by the structure of 
 a single chain, weighted by the appropriate chain length distribution.  Specifically, we have
 
 \begin{equation}
S(q) \equiv  \langle \frac{1}{N} \sum_{i,j} e^{-i \vec q \cdot (\vec r_i - \vec r_j)}\rangle,
\end{equation}
where $r_i$ is the coordinate of particle $i$, the sum runs over all
$N$ particles in the system, and the average $\langle...\rangle$ is over equilibrium configurations.
In the ideal gas limit, correlations between different chains can be neglected and $S(q)$ can be formally written as,
\begin{eqnarray}
S(q) =\frac{ \sum_{l=1}^{\infty} \rho_l l S_l(q)}{\sum_{l=1}^{\infty} \rho_l l} 
\label{eq:sq} 
\end{eqnarray}
where $S_l(q)$ is the structure factor (form factor) of a chain of length $l$:
\begin{eqnarray}
S_l(q)=\frac{1}{l} \langle \sum_{i,j=1}^l e^{-i \vec q \cdot (\vec r_i - \vec r_j)}\rangle
\end{eqnarray}

\begin{figure}[h]
\includegraphics[width=8cm, clip=true]{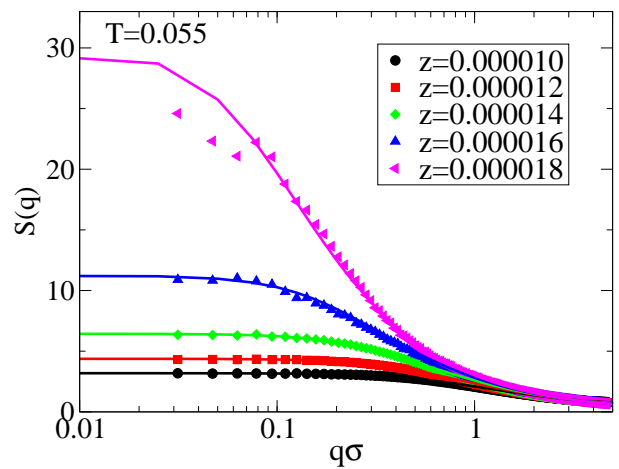}
\caption{Comparison of the structure factor calculated from the simulated configurations and from Eq.~\ref{eq:sq}-\ref{eq:sql} at T=0.055 for different values of the  activity.  }
\label{fig:sq}
\end{figure}

 Since the persistence length of the chains is $\approx 10-20$ particles  (see Fig.~\ref{fig:endtoend}), one can assume, as a first approximation, that in the investigated $T$-$\rho$ region chains are linear. When this is the case, averaging over all possible orientation of the chain gives:
 \begin{equation}
S_l(q)=1+\frac{1}{l}\sum_{j=1}^{l-1} 2 (l-j) \frac{sin(jq\sigma)}{jq\sigma}.
\label{eq:sql}
\end{equation}
The small $q$ expansion  of $S_l(q)$ is
\begin{equation}
S_l(q) \approx  l -\frac{l(l^2-1)}{36} (q \sigma)^2
\end{equation}
Correspondingly, $S(q)$ behaves at small $q$ as
\begin{eqnarray}
S(q) \approx \frac{<l^2>}{<l>} \left [ 1- \frac{q^2 \sigma^2}{36} \left  ( \frac{<l^4>-<l^2>}{<l^2>} \right )  \right ],
\end{eqnarray}
where $<l^m> \equiv \sum_l l^m \rho_l$ denotes the $m$ moment of the  cluster size distribution $\rho_l$.
Fig.~\ref{fig:sq} shows a comparison between the $S(q)$ calculated in the simulation and the theoretical $S(q)$ evaluated according to
Eq.~\ref{eq:sq} and \ref{eq:sql} at a low $T$, where the ideal
gas approximation is valid. Deviations are only observed at the highest density, suggesting that the ideal gas of chains is a good
representation of the structure of the system, in agreement with the
equivalence between  activity and monomer density shown in Fig.~\ref{fig:activity}. This observation is particularly relevant, since it suggests that  (in the appropriate $T$-$\rho$ window) also a description of the dynamics of the model based on the assumption of independent chains can be attempted.

\section{Conclusions}

In our pursuit of a fully predictive molecularly-based theory of self-assembly in terms of molecular parameters and liquid state correlation functions, we have considered a direct comparison of a liquid in which fluid particles have sticky spots on their polar regions to the predictions of the Wertheim theory for the relevant properties governing the  self-assembly thermodynamics. In the investigated region of temperatures and densities (which spans the  polymerization transition region), the predictions of Wertheim theory 
describe  simulation data remarkably well (without the use of any free parameters in these comparisons). This success means that we can be confident in pursuing more complicated types of self-assembly based on the foundation of Wertheim theory. For example, it is  possible to extend the Wertheim theory analysis in a direct way to describe the self-assembly of branched chains having multifunctional rather than ÔdipolarÕ symmetry interactions as in the present paper and to compare these results in a parameter free fashion to corresponding MC simulations for the these multifunctional interaction particles~\cite{LaNavepreprint}. Recent work has shown that the Wertheim theory describes the critical properties of these mutifunctional interaction particles rather well~\cite{bian}. According to this recent study, liquid phases of vanishing density can be generated once small fraction of polyfunctional particles are added to chain-forming models like the one studied here.  With the new generation of non-spherical sticky colloids, it should be possible to realize "empty liquids"~\cite{bian} and observe equilibrium gelation~\cite{genova,Zacca1}, i.e. approach dynamical arrest under equilibrium conditions.

The Werhtheim theory has also been applied successfully  to description of molecular associated liquids~\cite{Kol_87,Monson_98,simone} and to the thermodynamics of hard sphere polymer chains with short range attractive interactions~\cite{VegMac,GilVil}.
Thus, the theory could be adapted to describing mutually associating polymers and the formation of thermally reversible gels in these fluids upon cooling.

In summary, the Wertheim theory provides a promising framework for treating the thermodynamics of a wide range of self-assembling systems. The development of this theory and its validation by simulation and measurement should provide valuable tools in the practical development of self-assembly as a practical means of synthetic manufacturing. This theory also offers the prospect of improving the existing equilibrium association theories that are largely based on a lattice fluid model framework. This could allow progress to be made more rapidly, since this type of computation often offers computational advantages and because many problems such as chemically initiated chain branching~\cite {Tanaka}
 and thermally activated assembly processes have already been considered by lattice approaches~\cite {Dudo_03,Dudo_00}. 

 The problem of estimating the entropy of association in real self-assembling molecular and particle systems in solution is a difficult problem that has been addressed by many authors previously (Ref.~\cite{entropyassemblyreal} and refs. therein). It would clearly be interesting to extend the present work to determine how well the Wertheim theory could predict entropies of association for self-assembly processes that occur in a solvent rather than in the gas phase. The most interesting solvent in this connection, water, is a particular challenge since water itself can be considered an associating fluid, so that we are confronted with the problem of how the water association couples to the particle self-assembly. The problem of understanding the common tendency of particle self-assembly in aqueous solutions to occur upon heating  requires particular investigation. In the future, we look forward to exploring these more complex mixtures of associating fluids, which are so prevalent in real biological systems and in a materials processing context.  
 
As a final comment, we note that numerical work on this class of simple models  (playing with the particle interaction symmetries) can help understanding more complicated
ordered structures (as for example sheet-like, nanotube and closed nanoshell structures), as recently found when particles have multipole interaction potentials~\cite {Workum_06,Glotz_04,Glotz_Solomon,doye}.
We also note that the results discussed here apply to the growing field
of functionalized colloidal particles, colloidal particles with specifically designed shapes and interaction sites~\cite  {Manoh_03, Cho_05,Yi_04,Cho_05bis, Zerro_05,dna}. 

\section{Acknowledgments}

We acknowledge support from MIUR-Prin and
MCRTN-CT-2003-504712.  

\bibliographystyle{apsrev}
\bibliography{biblio_patchy}

\end{document}